## Research Article

# Efficient Algorithms and Implementation of a Semiparametric Joint Model for Longitudinal and Competing Risk Data: With Applications to Massive Biobank Data

Shanpeng Li,[1] Ning Li,[2,3] Hong Wang,[4] Jin Zhou,[1,2] Hua Zhou,[1,3] and Gang Li[1,3]

[1]*Department of Biostatistics, University of California at Los Angeles, Los Angeles, CA, USA*
[2]*Department of Medicine, University of California at Los Angeles, Los Angeles, CA, USA*
[3]*Department of Computational Medicine, University of California at Los Angeles, Los Angeles, CA, USA*
[4]*School of Mathematics and Statistics, Central South University, Changsha, China*

Correspondence should be addressed to Gang Li; vli@ucla.edu

Semiparametric joint models of longitudinal and competing risk data are computationally costly, and their current implementations do not scale well to massive biobank data. This paper identifies and addresses some key computational barriers in a semiparametric joint model for longitudinal and competing risk survival data. By developing and implementing customized linear scan algorithms, we reduce the computational complexities from $O(n^2)$ or $O(n^3)$ to $O(n)$ in various steps including numerical integration, risk set calculation, and standard error estimation, where $n$ is the number of subjects. Using both simulated and real-world biobank data, we demonstrate that these linear scan algorithms can speed up the existing methods by a factor of up to hundreds of thousands when $n > 10^4$, often reducing the runtime from days to minutes. We have developed an R package, FastJM, based on the proposed algorithms for joint modeling of longitudinal and competing risk time-to-event data and made it publicly available on the Comprehensive R Archive Network (CRAN).

## 1. Introduction

In clinical research and other longitudinal studies, it is common to collect both longitudinal and time-to-event data on each participant, and these two endpoints are often correlated. Joint models of longitudinal and survival data have been widely used to mitigate incorrect estimation and statistical inferences associated with separate analysis of each endpoint [1, 2]. For instance, in longitudinal data analysis, joint models are often used to handle *nonignorable* missing data due to a terminal event, which cannot be properly accounted for by a standard mixed effects model or generalized estimating equation (GEE) method that relies on the *ignorable* missing-at-random or missing-completely-at-random assumption [3–5]. Joint models are also popularly employed in survival analysis to study the effects of a time-dependent covariate that is measured intermittently or subject to measurement error [6–9] and for dynamic prediction of an event outcome from the past history of a biomarker [10–14]. Comprehensive reviews of joint models for longitudinal and time-to-event data and their applications can be found in Elashoff et al. [1], Hickey et al. [15], Rizopoulos [16], Sudell et al. [17] and the references therein.

Despite the explosive growth of literature on joint models for longitudinal and time-to-event data during the past three decades, efficient implementation of joint models has lagged behind, which limits the application of joint models to only small to moderate studies. Recently, massive sample size data collected from electronic health records (EHRs) and insurance claim databases warrants great opportunities to conduct clinical studies in a real-world setting. For example, the UK Biobank is a prospective cohort study with approximately 500,000 individuals, aged 37-73 years, from the general population between 2006 and 2010 in the United Kingdom [18, 19]. Aggregated and quality controlled EHR data purchasable from Optum (https://www.optum.com/EHR) includes 80+ millions patients with longitudinal lab measures. The Million Veteran Project [20] and IBM MarketScan Database [21] are some of many other big biobank databases that contain rich yet complex longitudinal and time-to-event data on $10^5 \sim 10^8$ patients. However, current implementations of many joint models are inefficient and not scalable to the size of biobank data as demonstrated later in Sections 3 and 4. There is a pressing need to develop efficient implementations of joint models to enable the analysis of these rich data sources.

The purpose of this paper is to develop and implement efficient algorithms for a semiparametric joint model of longitudinal and competing risk time-to-event data. As specified in Section 2.1, the joint model consists of a linear mixed effects submodel for a longitudinal outcome and a semiparametric proportional cause-specific hazard submodel for a competing risk survival outcome. These two submodels are linked together by shared random effects or features of an individual's longitudinal biomarker trajectory. In Section 2, we identify key computational bottlenecks in the semiparametric maximum likelihood inference procedure for the joint model. Specifically, we point out that in a standard implementation, the computational complexities for numerical integration, risk set calculation, and standard error estimation are of the order $O(n^2)$, $O(n^2)$, and $O(n^3)$, respectively, where $n$ is the number of subjects. Consequently, current implementation grinds to a halt as $n$ becomes large (for example, $n > 10^4$). We further develop tailored linear scan algorithms to reduce the computational complexity to $O(n)$ in each of the aforementioned components. We illustrate by simulation and real-world data that the linearization algorithms can result in a drastic speed-up by a factor of many thousands when $n > 10^4$, reducing the runtime from days to minutes for big data. Finally, we have developed a user-friendly R package FastJM to fit the shared parameter semiparametric joint model using the proposed efficient algorithms and made it publicly available on the Comprehensive R Archive Network (CRAN) at https://CRAN.R-project.org/package=FastJM.

The rest of the paper is organized as follows. Section 2.1 outlines the semiparametric shared random effect joint model framework and reviews a customized expectation-maximization (EM) algorithm for semiparametric maximum likelihood estimation as well as a standard error estimation method. Section 2.2 develops various linear scan algorithms to address the key computational bottlenecks in the EM algorithm and standard error estimation for large data. Section 3 presents simulation studies to illustrate the computational efficiency of the proposed linear scan algorithms. Section 4 demonstrates the improved computational performance of our FastJM R package over some established joint model R packages on two moderate to large real-world data sets. Concluding remarks are provided in Section 5.

## 2. Efficient Algorithms for a Semiparametric Joint Model of Longitudinal and Competing Risk Data

### 2.1. Notations and Preliminaries.
Let $Y_i(t)$ be the longitudinal outcome at time $t$ for subject $i$, $i = 1, 2, \cdots, n$, and $n$ is the total number of subjects. Suppose the longitudinal outcome $Y_i(t)$ is observed at time points $t_{ij}$, $j = 1, 2, \cdots, n_i$, and denote $Y_i = (Y_{i1}, \cdots, Y_{in_i})$. Let $C_i = (T_i, D_i)$ be the competing risk data on subject $i$, where $T_i$ is the observed time to either failure or censoring and $D_i$ takes value in $\{0, 1, 2, \cdots, K\}$, with $D_i = 0$ indicating a censored event and $D_i = k$ implying the $k$th type of failure is observed on subject $i$, $k = 1, 2, \cdots, K$. The censoring mechanism is assumed to be independent of the failure time.

#### 2.1.1. Model.
Consider the following joint model in which the longitudinal outcome $Y_i(t)$ is characterized by a linear mixed effects model:

$$Y_i(t) = m_i(t) + \varepsilon_i(t), = X_i^{(1)}(t)^T \beta + \tilde{X}_i^{(1)}(t)^T b_i + \varepsilon_i(t), \quad (1)$$

and the competing risk survival outcome is modeled by a proportional cause-specific hazard model:

$$\begin{aligned} \lambda_k\left(t \mid X_i^{(2)}, b_i, \gamma_k, \nu_k\right) &= \lim_{h \longrightarrow 0} \frac{P\left(t \leq T_i < t+h, D_i = k \mid T_i \geq t, X_i^{(2)}, b_i\right)}{h} \\ &= \lambda_{0k}(t) \exp\left\{X_i^{(2)T} \gamma_k + \nu_k^T b_i\right\}, k = 1, \cdots, K. \end{aligned} \quad (2)$$

In the longitudinal submodel (1), $m_i(t)$ is the mean of the longitudinal outcome at time $t$, $X_i^{(1)}(t)$ and $\tilde{X}_i^{(1)}(t)$ are column vectors of possibly time-varying covariates associated with the longitudinal outcome $Y_i(t)$, $\beta$ represents a $p \times 1$ vector of fixed effects of $X_i^{(1)}(t)$, $b_i \sim N_q(0, \Sigma)$ denotes a $q \times 1$ vector of random effects for $\tilde{X}_i^{(1)}(t)$, $\varepsilon_i(t) \sim N(0, \sigma^2)$ is the measurement error independent of $b_i$, and $\varepsilon_i(t_1)$ is independent of $\varepsilon_i(t_2)$ for any $t_1 \neq t_2$. In the competing risk survival submodel (2), $\lambda_k(t \mid X_i^{(2)}, b_i, \gamma_k, \nu_k)$ is the conditional cause-specific hazard rate for type $k$ failure at time $t$, given time-invariant covariates $X_i^{(2)}$ and the shared random effects $b_i$, and $\lambda_{0k}(t)$ is a completely unspecified baseline cause-specific hazard function for type $k$ failure. The two submodels are linked together by the shared random effects $b_i$, and the strength of association is quantified by the association parameters $\nu_1, \cdots, \nu_K$.

#### 2.1.2. Semiparametric Maximum Likelihood Estimation via an EM Algorithm.
Let $\Psi = \{\beta, \sigma^2, \gamma, \nu, \Sigma, \lambda_{01}(\cdot), \cdots, \lambda_{0K}(\cdot)\}$ denote all unknown parameters for the joint models (1) and (2), where $\gamma = (\gamma_1^T, \cdots, \gamma_K^T)^T$ and $\nu = (\nu_1^T, \cdots, \nu_K^T)^T$. Let $I(D_i = k)$ be the competing event indicator for type $k$ failure, which takes the value 1 if the condition $D_i = k$ is satisfied and 0 otherwise. The observed-data likelihood for $\Psi$ is then given by

$$L(\Psi;Y,C) \propto \prod_{i=1}^{n} f(Y_i, C_i \mid \Psi)$$

$$= \prod_{i=1}^{n} \int_{b_i} f(Y_i \mid C_i, b_i, \Psi) f(C_i \mid b_i, \Psi) f(b_i \mid \Psi) db_i \propto \prod_{i=1}^{n} \int_{b_i} \prod_{j=1}^{n_i} \frac{1}{\sqrt{2\pi\sigma^2}} \exp$$

$$\times \left[ -\frac{1}{2\sigma^2} \left\{ Y_{ij} - X_i^{(1)}(t_{ij})^T \beta - \tilde{X}_i^{(1)}(t_{ij})^T b_i \right\}^2 \right]$$

$$\times \left\{ \prod_{k=1}^{K} \lambda_k \left( t \mid X_i^{(2)}, b_i, \gamma_k, \nu_k \right)^{I(D_i=k)} \right\} \exp$$

$$\times \left[ -\int_0^{T_i} \left\{ \sum_{k=1}^{K} \lambda_k \left( t \mid X_i^{(2)}, b_i, \gamma_k, \nu_k \right) \right\} dt \right]$$

$$\times \frac{1}{\sqrt{(2\pi)^q |\Sigma|}} \exp\left( -\frac{1}{2} b_i^T \Sigma^{-1} b_i \right) db_i, \tag{3}$$

which follows from the assumption that $Y_i$ and $C_i$ are independent conditional on the covariates and the random effects.

Because $\Psi$ involves $K$ unknown hazard functions, finding its maximum likelihood estimate by maximizing the above observed-data likelihood is nontrivial. However, a customized EM algorithm can be derived to compute the maximum likelihood estimate of $\Psi$ by regarding the latent random effects $b_i$ as missing data [3]. The complete-data likelihood based on $(Y, C, b)$ is

$$L(\Psi; Y, C, b) \propto \prod_{i=1}^{n} \prod_{j=1}^{n_i} \frac{1}{\sqrt{2\pi\sigma^2}} \exp$$

$$\times \left[ -\frac{1}{2\sigma^2} \left\{ Y_{ij} - X_i^{(1)}(t_{ij})^T \beta - \tilde{X}_i^{(1)}(t_{ij})^T b_i \right\}^2 \right]$$

$$\times \prod_{k=1}^{K} \left\{ \Delta \Lambda_{0k}(T_i) \exp\left( X_i^{(2)T} \gamma_k + \nu_k^T b_i \right) \right\}^{I(D_i=k)} \tag{4}$$

$$\times \exp\left\{ -\sum_{k=1}^{K} \Lambda_{0k}(T_i) \exp\left( X_i^{(2)T} \gamma_k + \nu_k^T b_i \right) \right\}$$

$$\times \frac{1}{\sqrt{(2\pi)^q |\Sigma|}} \exp\left( -\frac{1}{2} b_i^T \Sigma^{-1} b_i \right),$$

where $\Lambda_{0k}(.)$ is the cumulative baseline hazard function for type $k$ failure and $\Delta\Lambda_{0k}(T_i) = \Lambda_{0k}(T_i) - \Lambda_{0k}(T_i-)$.

The EM algorithm iterates between an expectation step (E-step):

$$Q\left(\Psi; \Psi^{(m)}\right) \equiv E^{(m)}_{b|Y,C,\Psi^{(m)}} \{\log L(\Psi; Y, C, b)\}, \tag{5}$$

and a maximization step (M-step):

$$\Psi^{(m+1)} = \arg \max_{\Psi} Q\left(\Psi; \Psi^{(m)}\right), \tag{6}$$

until the algorithm converges, where $\Psi^{(m)}$ is the estimate of $\Psi$ from the $m$th iteration. The E-step (5) involves calculating the following expected values across all subjects: $E^{(m)}(b_i)$, $E^{(m)}(b_i b_i^T)$, $E^{(m)}\{\exp(\nu_k b_i)\}$, $E^{(m)}\{b_i \exp(\nu_k b_i)\}$, and $E^{(m)}\{b_i b_i^T \exp(\nu_k b_i)\}$, where

$$E^{(m)}\{h(b_i)\} = \int h(b_i) f\left(b_i \mid Y_i, C_i, \Psi^{(m)}\right) db_i, \tag{7}$$

for any function $h(\cdot)$. Furthermore, it can be shown that the M-step (6) has closed-form solutions for the parameters $\beta$, $\sigma^2$, $\Sigma$, and $\Lambda_{0k}(t)$, and that the other parameters $\gamma$ and $\nu$ can be updated using the one-step Newton-Raphson method. Details are provided in equations (7.1)–(7.8) of the supplementary materials (available here).

*2.1.3. Standard Error Estimation.* As discussed in Elashoff et al. [22] (Section 4.1, p.72), standard errors of the parametric components of the semiparametric maximum likelihood estimate can be estimated by profiled likelihood, observed information matrix, or bootstrap. All three methods can be computationally intensive when $n$ is large. Here, we focus on the profiled likelihood-based method and show that its computation can be linearized with respect to $n$.

Let $\Omega = (\beta, \Sigma, \sigma^2, \gamma_1, \cdots, \gamma_K, \nu_1, \cdots, \nu_K)$ denote the parametric component of $\Psi$ and $\hat{\Omega}$ its maximum likelihood estimate. The variance-covariance matrix of $\hat{\Omega}$ can be estimated by inverting the following approximate empirical Fisher information [23–25]:

$$\sum_{i=1}^{n} \nabla_{\Omega} l^{(i)}\left(\hat{\Omega}; Y, C\right) \nabla_{\Omega} l^{(i)}\left(\hat{\Omega}; Y, C\right)^T, \tag{8}$$

where $\nabla_{\Omega} l^{(i)}(\Omega; Y, C)$ is the observed score vector from the profiled likelihood $l^{(i)}(\Omega; Y, C)$ of $\Omega$ on the $i$th subject by profiling out the baseline hazards. Details of the observed score vector for each parametric component are provided in equations (7.9)–(7.13) of the supplementary materials.

*2.2. Efficient Algorithms and Implementation of the EM Algorithm and Standard Error Estimation.* With naive implementation, multiple quantities in the above E-step, M-step, and standard error estimation will involve $O(n^2)$ or $O(n^3)$ operations, which become computationally prohibitive at large sample size $n$. Below we identify these bottlenecks and discuss appropriate linear scan algorithms to reduce their computational complexities to $O(n)$.

*2.2.1. Efficient Implementation of the E-Step.* At each EM iteration, the E-step (5) requires calculating integral (7) across all subjects. Below we discuss how to accelerate two commonly used numerical integration methods for evaluating these integrals.

*(1) Standard Gauss-Hermite Quadrature Rule for Numerical Integration.* A commonly used method for numerical evaluation of integral (7) is based on the standard Gauss-Hermite quadrature rule [26]:

$$E^{(m)}\{h(b_i)\} = \frac{\int h(b_i) f(Y_i, C_i, b_i \mid \Psi^{(m)}) db_i}{f(Y_i, C_i \mid \Psi^{(m)})}$$

$$= \frac{\int h(b_i) f(Y_i, C_i \mid b_i, \Psi^{(m)}) f(b_i \mid \Psi^{(m)}) db_i}{\int f(Y_i, C_i \mid b_i, \Psi^{(m)}) f(b_i \mid \Psi^{(m)}) db_i}$$

$$\approx \frac{\sum_{t_1, t_2, \cdots, t_q} \pi_t h(\tilde{b}_t^{(m)}) f(Y_i, C_i \mid \tilde{b}_t^{(m)}, \Psi^{(m)}) f(\tilde{b}_t^{(m)} \mid \Psi^{(m)}) \exp(\|c_t\|^2)}{\sum_{t_1, t_2, \cdots, t_q} \pi_t f(Y_i, C_i \mid \tilde{b}_t^{(m)}, \Psi^{(m)}) f(\tilde{b}_t^{(m)} \mid \Psi^{(m)}) \exp(\|c_t\|^2)},$$

(9)

where

$$f(Y_i, C_i \mid \tilde{b}_t^{(m)}, \Psi^{(m)}) = f(Y_i \mid \tilde{b}_t^{(m)}, \Psi_y^{(m)}) \times f(C_i \mid \tilde{b}_t^{(m)}, \Psi_c^{(m)})$$

$$= \prod_{j=1}^{n_i} \frac{1}{\sqrt{2\pi\sigma^{2(m)}}} \exp$$

$$\cdot \left[ -\frac{1}{2\sigma^{2(m)}} \left\{ Y_{ij} - X_i^{(1)}(t_{ij})^T \beta^{(m)} - \tilde{X}_i^{(1)}(t_{ij})^T \tilde{b}_t^{(m)} \right\}^2 \right]$$

$$\times \prod_{k=1}^{K} \left\{ \Delta \Lambda_{0k}^{(m)}(T_i) \exp\left( X_i^{(2)T} \gamma_k^{(m)} + v_k^{T(m)} \tilde{b}_t^{(m)} \right) \right\}^{I(D_i = k)}$$

$$\times \exp \left\{ -\sum_{k=1}^{K} \Lambda_{0k}^{(m)}(T_i) \exp\left( X_i^{(2)T} \gamma_k^{(m)} + v_k^{T(m)} \tilde{b}_t^{(m)} \right) \right\},$$

(10)

with $\Lambda_{0k}^{(m)}(.)$ the right-continuous and nondecreasing cumulative baseline hazard function for type $k$ failure at the $m$th EM iteration as defined in Section 7.1 (equation (7.4)) of the supplementary material and $\Delta \Lambda_{0k}^{(m)}(T_i)$ the jump size of $\Lambda_{0k}^{(m)}(.)$ at $T_i$, $q$ the dimension of the random effects vector, $\sum_{t_1, t_2, \cdots, t_q}$ the shorthand for $\sum_{t_1=1}^{n_q} \cdots \sum_{t_q=1}^{n_q}$, $n_q$ the number of quadrature points, $c_t = (c_{t_1}, c_{t_2}, \cdots, c_{t_q})^T$ the abscissas with corresponding weights $\pi_t$, $\tilde{b}_t^{(m)} = \sqrt{2} \Sigma \Lambda^{(m)1/2} c_t$ the rescaled alternative abscissas, and $\Sigma \Lambda^{(m)1/2}$ the square root of $\Sigma \Lambda^{(m)}$ [3]. However, this method is computationally intensive due to multiple factors. First, it usually requires many quadrature points to approximate an integral with sufficient accuracy because the mode of the integrand is often located in a region different from zero. Second, the computational cost increases exponentially with $q$ because the Cartesian product of the abscissas is used to evaluate the integrand with respect to each random effect. Lastly, the alternative abscissas $\tilde{b}_t^{(m)}$ need to be recalculated at every EM iteration.

*(2) Pseudo-Adaptive Gauss-Hermite Quadrature Rule for Numerical Integration.* When the number of longitudinal measurements per subject is relatively large, Rizopoulos [27] introduced a pseudo-adaptive Gauss-Hermite quadrature rule for numerical approximation of integral (7), which achieves good approximation accuracy with only a small number ($n_q$) of quadrature points and is thus computationally more efficient. The pseudo-adaptive Gauss-Hermite quadrature rule proceeds as follows. First, fit the linear mixed effects model (1) to extract the empirical Bayes estimates of the random effects and its covariance matrix $\tilde{H}_i^{-1}$:

$$\tilde{b}_i = \widehat{\Sigma}_i \tilde{X}_i^{(1)} V \Lambda_i^{-1} \left\{ Y_i - X_i^{(1)T} \widehat{\beta} \right\},$$  (11)

$$\tilde{H}_i^{-1} = \widehat{\Sigma} - \widehat{\Sigma}_i \tilde{X}_i^{(1)} \left[ V \Lambda_i^{-1} - V \Lambda_i^{-1} X_i^{(1)T} \left\{ \sum_{i=1}^{n} X_i^{(1)} V \Lambda_i^{-1} X_i^{(1)T} \right\}^{-1} X_i^{(1)} V \Lambda_i^{-1} \right]$$

$$\times \tilde{X}_i^{(1)T} \widehat{\Sigma},$$

(12)

where $X_i^{(1)} = (X_i^{(1)}(t_{i1}), \cdots, X_i^{(1)}(t_{in_i}))$, $\tilde{X}_i^{(1)} = (\tilde{X}_i^{(1)}(t_{i1}), \cdots, \tilde{X}_i^{(1)}(t_{in_i}))$, and $V_i = {}_i^{(1)T} \Sigma_i^{(1)} + \sigma^2 I$. Then, define the alternative abscissas $\tilde{r}_t = \tilde{b}_i + \sqrt{2} \tilde{H}_i^{-1/2} c_t$ and approximate $E^{(m)}\{h(b_i)\}$ by

$$E^{(m)}\{h(b_i)\} \approx \frac{\sum_{t_1, t_2, \cdots, t_q} \pi_t h(\tilde{r}_t) f(Y_i, C_i \mid \tilde{r}_t, \Psi^{(m)}) f(\tilde{r}_t \mid \Psi^{(m)}) \exp\{\|c_t\|^2\}}{\sum_{t_1, t_2, \cdots, t_q} \pi_t f(Y_i, C_i \mid \tilde{r}_t, \Psi^{(m)}) f(\tilde{r}_t \mid \Psi^{(m)}) \exp\{\|c_t\|^2\}},$$

(13)

where

$$f(Y_i, C_i \mid \tilde{r}_t, \Psi^{(m)}) = f(Y_i \mid \tilde{r}_t, \Psi_y^{(m)}) \times f(C_i \mid \tilde{r}_t, \Psi_c^{(m)})$$

$$= \prod_{j=1}^{n_i} \frac{1}{\sqrt{2\pi\sigma^{2(m)}}} \exp$$

$$\cdot \left[ -\frac{1}{2\sigma^{2(m)}} \left\{ Y_{ij} - X_i^{(1)}(t_{ij})^T \beta^{(m)} - \tilde{X}_i^{(1)}(t_{ij})^T \tilde{r}_t \right\}^2 \right]$$

$$\times \prod_{k=1}^{K} \left\{ \Delta \Lambda_{0k}^{(m)}(T_i) \exp\left( X_i^{(2)T} \gamma_k^{(m)} + v_k^{(m)T} \tilde{r}_t \right) \right\}^{I(D_i = k)}$$

$$\times \exp \left\{ -\sum_{k=1}^{K} \Lambda_{0k}^{(m)}(T_i) \exp\left\{ X_i^{(2)T} \gamma_k^{(m)} + v_k^{(m)T} \tilde{r}_t \right\} \right\},$$

(14)

with the notations defined similarly to Equation (10).

A derivation of (13) is provided in Section 7.2 of the supplementary materials. The pseudo-adaptive Gauss-Hermite quadrature rule is computationally appealing because the alternate rescaled quadrature points $\tilde{r}_t$ are computed only once before the EM algorithm and do not need to be updated in the EM algorithm. Additionally, the pseudo-adaptive Gauss-Hermite quadrature rule requires fewer quadrature points than the standard Gauss-Hermite quadrature rule to achieve the same numerical approximation accuracy [27]. For example, our simulation results in the supplementary materials (Section A.4, Table A.1) illustrate that the pseudo-adaptive Gauss-Hermite quadrature rule with $n_q = 6$ quadrature points produces almost identical results to the standard Gauss-Hermite quadrature rule with $n_q = 20$ quadrature points.

*Remark 1* (Linear calculation of $\tilde{H}_i^{-1}$'s across all subjects). At the first sight, calculating $\tilde{H}_i^{-1}$ (12) across all subjects would involve $O(n^2)$ operations since each $\tilde{H}_i^{-1}$ involves a summation over $n$ subjects. However, because the same quantity $\sum_{i=1}^{n} X_i^{(1)} \widehat{V}_i^{-1} X_i^{(1)T}$ appears in every $\tilde{H}_i^{-1}$, one can precompute this quantity and then use the cached value to calculate $\tilde{H}_i^{-1}$ across all subjects. This way, one can compute the $\tilde{H}_i^{-1}$'s across all subjects in $O(n)$ operations. Our simulation study in the supplementary material (Section A.5, Figure A.1) shows that applying this simple linearization algorithm can yield a speed-up by a factor of 10 to 10,000 when $n$ grows from 10 to $10^5$. Our implementation is also significantly faster than a popular R package lme4 [28] with over a speed-up by a factor of 10 to 500 as $n$ grows from 10 to $10^5$. *(3) Linear Scan Algorithm for Calculating $f(Y_i, C_i | \tilde{r}_t, \Psi^{(m)})$ across All subjects.* Both the standard Gauss-Hermite quadrature rule (9) and the pseudo-adaptive Gauss-Hermite quadrature rule (13) require evaluating $f(Y_i, C_i | b_i, \Psi^{(m)})$ at their prespecified abscissas across all subjects (see Equations (10) and (14)). Hence, calculating $f(Y_i, C_i | \tilde{r}_t, \Psi^{(m)})$ requires evaluation of $\Lambda_{0k}^{(m)}(T_i)$ across all subjects for each $k$. We observe from equation (7.4) that for each $k$, $\{\Lambda_{0k}^{(m)}(t_{k1}), \cdots, \Lambda_{0k}^{(m)}(t_{kq_k})\}$ have already been calculated from the $m$th EM iteration, where $t_{k1} > \cdots > t_{kq_k}$ are $q_k$ distinct observed type $k$ event times. For each $i$, calculating $\Lambda_{0k}^{(m)}(T_i)$ would involve $O(n)$ operations if a global search is performed to find the interval of two adjacent type $k$ event times containing $T_i$. Consequently, calculating $\{\Lambda_{0k}^{(m)}(T_1), \cdots, \Lambda_{0k}^{(m)}(T_n)\}$ would require $O(n^2)$ operations. However, by taking advantage of the fact that $\Lambda_{0k}^{(m)}(t)$ is a right-continuous and nondecreasing step function, one can obtain $\{\Lambda_{0k}^{(m)}(T_1), \cdots, \Lambda_{0k}^{(m)}(T_n)\}$ from $\{\Lambda_{0k}^{(m)}(t_{k1}), \cdots, \Lambda_{0k}^{(m)}(t_{kq_k})\}$ in $O(n)$ operations using the following linear scan algorithm. First, sort the observation times $T_i$, $i = 1, \cdots, n$, in descending order. Denote by $(i)$ the ranked index of a subject. Then, define a mapping

$$\{\Lambda_{0k}^{(m)}(t_{k1}), \Lambda_{0k}^{(m)}(t_{k2}), \cdots \Lambda_{0k}^{(m)}(t_{kq_k})\} \mapsto \\ \cdot \{\Lambda_{0k}^{(m)}(T_{(1)}), \Lambda_{0k}^{(m)}(T_{(2)}), \cdots, \Lambda_{0k}^{(m)}(T_{(n)})\}, \quad (15)$$

where $t_{k1}, \cdots, t_{kq_k}$ are scanned forward from the largest to the smallest, and for each $t_{kj}$, only a subset of the ranked observation times $T_{(i)}$ are scanned forward to calculate $\Lambda_{0k}^{(m)}(T_{(i)})$ as follows:

$$\Lambda_{0k}^{(m)}(T_{(i)}) = \begin{cases} \Lambda_{0k}^{(m)}(t_{k1}), & \text{if } T_{(i)} \geq t_{k1}, \\ \Lambda_{0k}^{(m)}(t_{k(j+1)}), & \text{if } T_{(i)} \in [t_{k(j+1)}, t_{kj}), \text{for some } j \in \{1, \cdots, q_k - 1\}, \\ 0, & T_{(i)} < t_{kq_k}. \end{cases} \quad (16)$$

Specifically, start with $t_{k1}$ and scan through the first set of observation times $T_{(1)} \geq \cdots \geq T_{(i_{k1})}$ where $T_{(i_{k1})} \geq t_{k1} > T_{(i_{k1})+1}$, and the corresponding $\Lambda_{0k}(T_{(i)})$'s take the value $\Lambda_{0k}(t_{k1})$. Next, move forward to $t_{k2}$ and scan through the second set $T_{(i_{k1}+1)} \geq \cdots \geq T_{(i_{k2})}$, where $T_{(i_{k2})} \geq t_{k2} > T_{(i_{k2})+1}$, and the corresponding $\Lambda_{0k}(T_{(i)})$'s take the value evaluated at $\Lambda_{0k}(t_{k2})$. Repeat the same process until $T_{(n)}$ is scanned. Because the scanned $T_{(i)}$'s for different $t_{kj}$'s do not overlap, the entire algorithm costs only $O(n)$ operations.

*2.2.2. Linear Risk Set Scan for the M-Step.* In the M-step, multiple quantities in equations (A.4)–(A.8), such as the cumulative baseline hazard functions and the Hessian matrix and score vector for $\gamma_k$ and $\nu_k$ ($k = 1, 2, \cdots, K$), involve aggregating information over the risk set $R(t_{kj}) = \{r : T_r \geq t_{kj}\}$ at each uncensored event time $t_{kj}$. These quantities are further aggregated across all $t_{kj}$'s. If all subjects are scanned to determine the risk set $R(t_{kj})$ at each $t_{kj}$, then aggregating information over the risk set for all uncensored event times would obviously require $O(n^2)$ operations. Below we explain how to linearize the number of operations for risk set calculations across all uncensored event times by taking advantage of the fact that the risk set is decreasing over time for right censored data.

First, to calculate $\Lambda_{0k}^{(m+1)}(t_{kj})$, $j = 1, \cdots, q_k$, one needs to compute $\sum_{r \in R(t_{kj})} \exp(X_r^{(2)T} \gamma_k^{(m)}) E\{\exp(\nu_k^{(m)} b_r)\}$, $j = 1, \cdots, q_k$. Because the distinct uncensored event times $t_{k1} > \cdots > t_{kq_k}$ are arranged in decreasing order, the risk set $R(t_{k(j+1)})$ can be decomposed into two disjoint sets: $R(t_{k(j+1)}) = R(t_{kj}) \cup \{r : T_{(r)} \in [t_{k(j+1)}, t_{kj})\}$, and consequently,

$$\sum_{r \in R(t_{k(j+1)})} a_r = \sum_{r \in R(t_{kj})} a_r + \sum_{\{r : T_{(r)} \in [t_{k(j+1)}, t_{kj})\}} a_r, \quad (17)$$

for any sequence of real numbers $a_1, \cdots, a_n$. It follows from the recursive formula (17) and the fact that the subjects in $R(t_{kj})$ do not need to be scanned to calculate the second term of (17), one can calculate $\sum_{r \in R(t_{kj})} a_r$, $j = 1, \cdots q_k$, in $O(n)$ operations when $T_{(r)}$'s are scanned backward in time.

Next, to calculate the Hessian matrix $I_{\gamma_k}^{(m)}$ for $\gamma_k$ in (A.5), we first rewrite it as

$$I_{\gamma_k}^{(m)} = \sum_{j=1}^{q_k} \left[ \Delta\Lambda_{0k}(t_{kj})^{(m+1)} \sum_{r \in R(t_{kj})} \exp(X_r^{(2)T} \gamma_k^{(m)}) E \right. \\ \left. \cdot \{\exp(\nu_k^{(m)T} b_r)\} X_r^{(2)} X_r^{(2)T} \right], \quad (18)$$

which allows one to linearize its calculation based on (17) with $a_r = \Delta\Lambda_{0k}(t_{kj})^{(m+1)} \exp(X_r^{(2)T} \gamma_k^{(m)}) E\{\exp(\nu_k^{(m)T} b_r)\} X_r^{(2)} X_r^{(2)T}$ similar to the linear scan algorithm for $\Lambda_{0k}^{(m+1)}(t_{kj})$'s.

Finally, the above linear risk set scan algorithm can be adapted to calculate the Hessian matrix and score vector for $\gamma_k$ and $\nu_k$ in equations (A.6)-(A.8) in $O(n)$ operations in a similar fashion.

*2.2.3. Linear Scan Algorithm for Standard Error Estimation.*
The standard error estimation formula in (8) relies on the observed score vector from the profiled likelihood where the baseline hazard is profiled out. However, for each subject $i$, two components of the score vector, $\nabla_{\gamma_k} l^{(i)}(\hat{\Omega}; Y, C)$ and $\nabla_{\nu_k} l^{(i)}(\hat{\Omega}; Y, C)$ as given in equations (A.12) and (A.13), involve aggregating information either over $\{r \in R(T_i)\}$ or over both $\{r \in R(t_{kj})\}$ and $\{j : t_{kj} \leq T_i\}$. If implemented naively, the aggregation can take either $O(n)$ or $O(n^2)$ operations, respectively. As a result, the observed information matrix can take $O(n^3)$ operation as it requires summing up the information across all subjects. Below we describe a sequential linear scan algorithm to reduce the computational complexity from $O(n^3)$ to $O(n)$.

Our algorithm can be easily explained by considering the calculation of the following expression in the second term of $\nabla_{\gamma_k} l^{(i)}(\hat{\Omega}; Y, C)$ in (7.12):

$$\sum_{j:t_{kj} \leq T_i} \frac{d_{kj} \sum_{r \in R(t_{kj})} \exp\left(\gamma_k^T X_r^{(2)}\right) E\{\exp\left(\nu_k^T b_r\right)\} X_r^{(2)}}{\left[\sum_{r \in R(t_{kj})} \exp\left(\gamma_k^T X_r^{(2)}\right) E\{\exp\left(\nu_k^T b_r\right)\}\right]^2}, \text{ for } i = 1, \cdots, n. \tag{19}$$

In other words, we need to compute $B(T_i)$ for $i = 1, \cdots, n$, where $B(t) \equiv \sum_{j:t_{kj} \leq t} b_{kj}$ and

$$b_{kj} = \frac{d_{kj} \sum_{r \in R(t_{kj})} \exp\left(\gamma_k^T X_r^{(2)}\right) E\{\exp\left(\nu_k^T b_r\right)\} X_r^{(2)}}{\left[\sum_{r \in R(t_{kj})} \exp\left(\gamma_k^T X_r^{(2)}\right) E\{\exp\left(\nu_k^T b_r\right)\}\right]^2}. \tag{20}$$

Before going further, we recall that the distinct uncensored event times $t_{k1} > \cdots > t_{kq_k}$ are in descending order and that the subjects are sorted so that the observation times $T_i$'s are in descending order.

First of all, because the risk set is decreasing over time for right censored data, it follows from Equation (17) that $B(t_{k1}), \cdots, B(t_{kq_k})$ can be computed in $O(n)$ operations as one scans through $t_{k1}, \cdots, t_{kq_k}$ backward in time. Second, analogous to (15), the following linear scan algorithm can be used to calculate $\{B(T_{(1)}), B(T_{(2)}), \cdots, B(T_{(n)})\}$ from $\{B(t_{k1}), \cdots, B(t_{kq_k})\}$:

$$\left\{B(t_{k1}), \cdots, B\left(t_{kq_k}\right)\right\} \mapsto \left\{B\left(T_{(1)}\right), B\left(T_{(2)}\right), \cdots, B\left(T_{(n)}\right)\right\}, \tag{21}$$

where $t_{k1}, \cdots, t_{kq_k}$ are scanned forward from the largest to the smallest, and for each $t_{kj}$, only a subset of the ranked observation times $T_{(i)}$'s are scanned forward to calculate $B(T_{(i)})$'s as follows:

$$B\left(T_{(i)}\right) = \begin{cases} B(t_{k1}), & \text{if } T_{(i)} \geq t_{k1}, \\ B\left(t_{k(j+1)}\right), & \text{if } T_{(i)} \in \left[t_{k(j+1)}, t_{kj}\right), \text{ for some } j \in \{1, \cdots, q_k - 1\}, \\ 0, & \text{otherwise.} \end{cases} \tag{22}$$

The details are essentially the same as those discussed following Equation (15) and thus omitted here.

## 3. Simulation Studies

We present a simulation study to illustrate the computational speed-up rendered by the proposed linear algorithms as the sample size $n$ grows from 100 to 1,000,000. All simulations were run on a MacBook Pro with 6-Core Intel Core i7 processor (2.6 GHz) and 16 GB RAM running MacOS.

We generated longitudinal measurements $Y_{ij}$ from

$$Y_{ij} = \beta_0 + \beta_1 t_{ij} + \beta_2 X_{2i} + b_{0i} + b_{1i} t_{ij} + \varepsilon_{ij}, \tag{23}$$

which corresponds to model (1) with $X_i^{(1)}(t_{ij})^T = (1, t_{ij}, X_{2i})$ and $\tilde{X}_i^{(1)}(t_{ij})^T = (1, t_{ij})$, and competing risk event times from a proportional cause-specific hazard model

$$\lambda_1(t; X_{1i}, X_{2i}, b_i, \gamma_1, \nu_1) = \lambda_{01}(t) \exp\{\gamma_{11} X_{1i} + \gamma_{12} X_{2i} + \nu_1^T b_i\}, \tag{24}$$

$$\lambda_2(t; X_{1i}, X_{2i}, b_i, \gamma_2, \nu_2) = \lambda_{02}(t) \exp\{\gamma_{21} X_{1i} + \gamma_{22} X_{2i} + \nu_2^T b_i\}, \tag{25}$$

where the two submodels (23)–(25) are linked together through the shared random effects $b_i = (b_{0i}, b_{1i})^T$. In the above joint model, $t_{ij} = 0, 1, \cdots$ represent scheduled visit times, $X_{1i}$ follows $N(2, 1.0)$, $X_{2i} \sim \text{Bernoulli}(0.5)$ is a binary covariate, the random effects $b_i = (b_{0i}, b_{1i})^T$ follows a $N_2(0, \Sigma)$ distribution with $\Sigma_{11} = 0.5$, $\Sigma_{22} = 0.25$, and $\Sigma_{12} = 0$, the measurement errors $\varepsilon_{ij}$ are iid $N(0, 0.5)$ and independent of $b_i$, and the baseline hazards $\lambda_{01}(t)$ and $\lambda_{01}(t)$ are constants 0.05 and 0.1, respectively. We simulated noninformative censoring time $V_i$ following $\exp(20)$ and let $T_i = \min\{T_{i1}^*, T_{i2}^*, V_i\}$ be the observed time (possibly censored) for subject $i$. The longitudinal measurements for subject $i$ at $t_{ij}$ are assumed missing after $T_i$.

We first compared the runtime between three different implementations of the EM algorithm for fitting the joint models (1) and (2) as described in Section 2.2.

(1) *Method 1*. This method is a standard implementation of the EM algorithm using the standard Gauss-Hermite quadrature rule in the E-step (Equation (9) with $n_q = 20$) without any linear computation.

(2) *Method 2*. This is a standard implementation of the EM algorithm using the pseudo-adaptive quadrature rule in the E-step (Equation (13) with $n_q = 6$) with the linear calculation of $\tilde{H}_i^{-1}$'s described in Remark 1 and without any other linear computation.

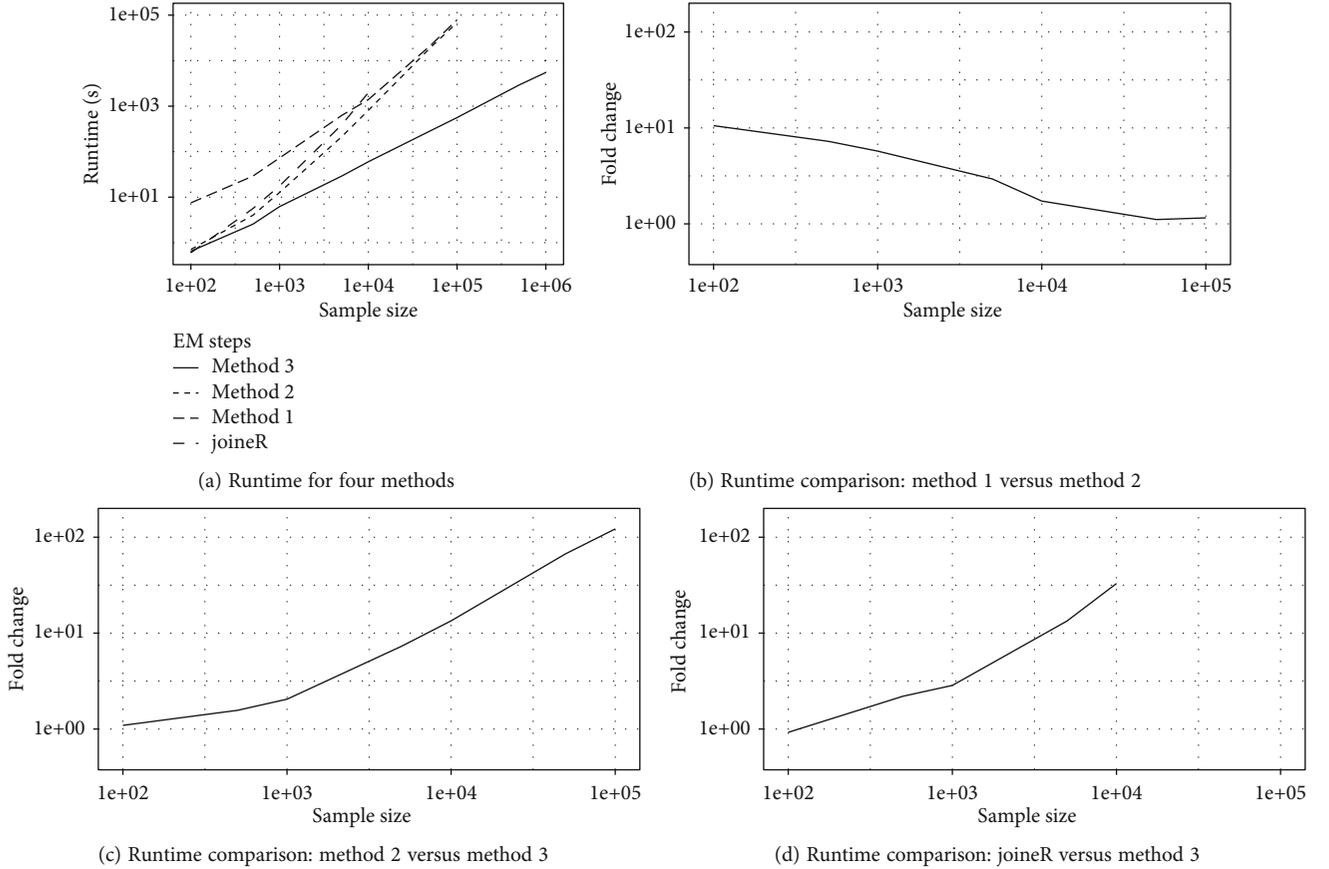

Figure 1: Runtime (seconds) comparison between three different implementations of the EM algorithm for fitting the joint models (1) and (2) and the joineR package. The details of methods 1-3 are given in Section 3. joineR is an established R package which fits a similar semiparametric joint model with a slightly different latent association structure in the competing risk submodel [29]. Fold change is calculated as the ratio of runtime between two methods.

(3) *Method 3*. This is a method 2+linear scan for calculating $f(Y_i, C_i \mid \tilde{r}_t, \Psi^{(m)})$'s+linear risk set scan for M-step as described in Section 2.2.

The number of quadrature points $n_q$ for methods 1 and 2 was determined by first trying different values, {10, 20, 30} for method 1 and {6, 9, 12, 15} for method 2, and then choosing the smallest value for which the estimation results are stabilized and similar between the two implementation methods. For comparison purposes, we have also included the runtime of an established joint model R package joineR, which uses a similar EM algorithm for parameter estimation to fit a semiparametric joint model with a slightly different latent association structure in the competing risk submodel [29]. The results are depicted in Figure 1.

It is seen from Figure 1(a) that the runtime of method 3 increases linearly with the sample size, while the runtime of the other three methods grows exponentially. For moderate sample size, method 2 is computationally more efficient than method 1 because it requires fewer quadrature points for numerical integration. However, its computational advantage diminishes as the sample size increases due to the exponentially increasing computational cost of $f(Y_i, C_i \mid \tilde{r}_t, \Psi^{(m)})$'s and risk set calculation in the M-step. By further linearizing the computation of these key components, method 3 has yielded more than 100-fold speed-up over method 2 when $n = 10^5$, and the speed-up is expected to increase exponentially as $n$ increases (Figure 1(b)). Furthermore, method 3 has demonstrated more than 30-fold speed-up over joineR when $n = 10^4$. We also note that joineR failed to run when $n = 10^5$ due to the overload of memory.

We also compared the runtime of two implementations of the standard error estimation, with and without linear scan as described in Section 2.2.3, and the bootstrap method employed by the joineR package [29]. The results are shown in Figure 2.

It is seen from Figure 2(a) that the implementation with linear scan easily scales to a million subjects, taking only minutes to finish, while the naive implementation without linear scan grinds to a halt when the sample size is 10,000 or larger. Figure 2(b) shows that linear scan can generate a speed-up by a factor of greater than 100,000 when $n \geq 10,000$. Similarly, in comparison with joineR that used 100 bootstrap samples for standard error estimation, our standard error estimation method with linear scan generated a speed-up by a factor of greater than 100,000 when $n \geq 5,000$.

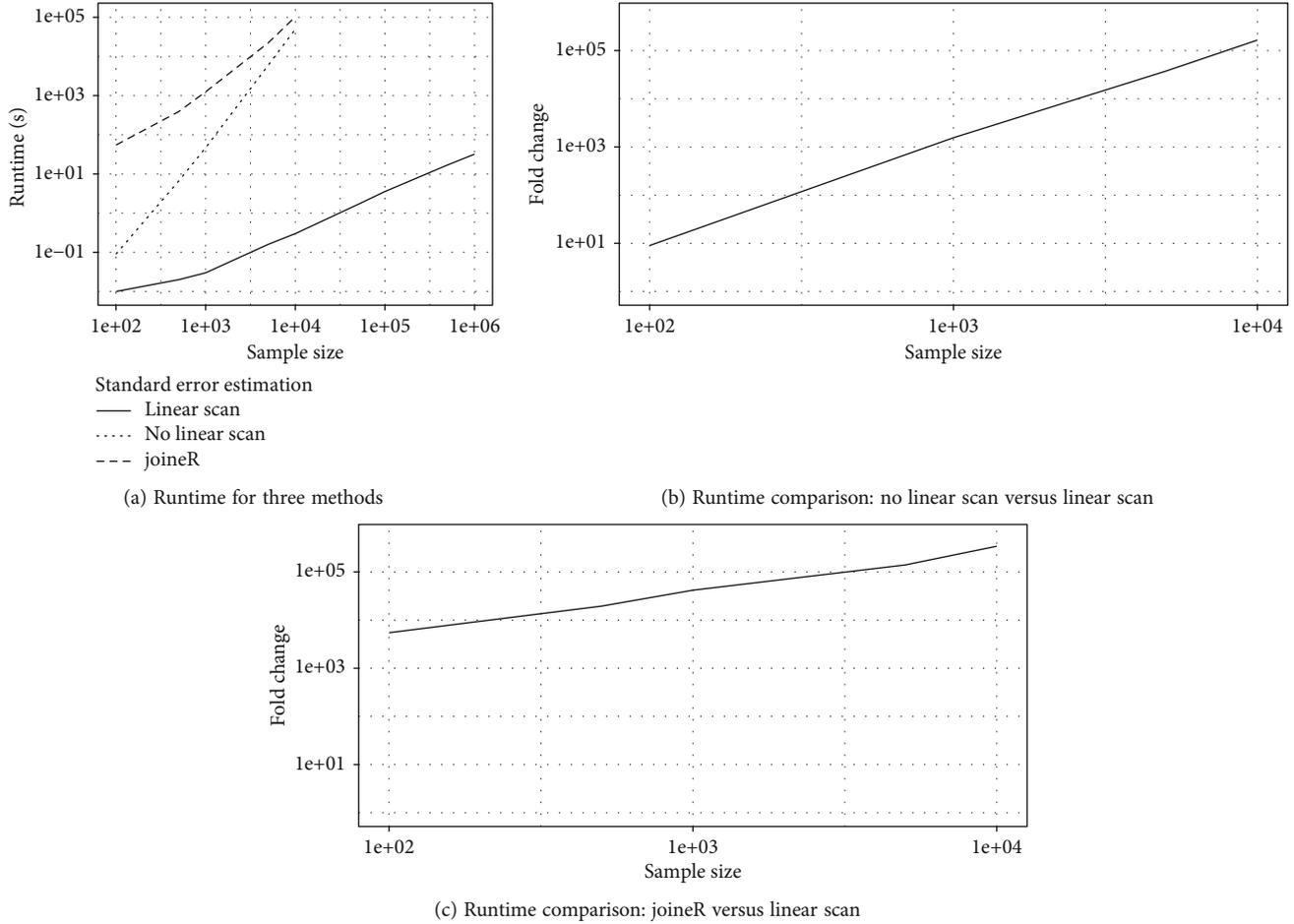

(a) Runtime for three methods

(b) Runtime comparison: no linear scan versus linear scan

(c) Runtime comparison: joineR versus linear scan

Figure 2: Runtime (seconds) comparison between two implementations of standard error estimation for fitting the joint models (1) and (2), linear scan and no linear scan as described in Section 2.2.3, and the bootstrap method employed by the joineR package [29]. Fold change is calculated as the ratio of runtime between two methods.

Finally, Figures 1 and 2 in Section 3 have focused on contrasting the computational efficiency of different implementations for parameter estimation and standard error estimation in terms of the runtime. We have also compared their parameter estimates and standard error in Section A.6 of the supplementary materials. As one would expect, our three different implementations (methods 1-3) yielded almost identical estimation results, whereas joineR produced similar estimation results for the longitudinal model, but slightly different results for the competing risk model due to its different latent association structure.

## 4. Real Data Examples

We have developed an R package FastJM to implement the efficient algorithms described in Section 2. Below we illustrate the improved computational performance of FastJM in comparison to existing joint model R packages on a lung health study (LHS) data with $n = 5,887$ subjects and a UK Biobank data with $n = 193,287$ participants.

*4.1. Lung Health Study.* The lung health study (LHS) data were collected from a ten-center randomized clinical trial on 5,887 middle-aged smokers with mild to moderate chronic obstructive pulmonary disease (COPD) [30]. Patients were randomized into three arms: usual care, smoking intervention and placebo inhaler (SIP), and smoking intervention and active bronchodilator inhaler (SIA). An important objective of the study was to determine if the intervention program with the combination of intensive smoking cessation counseling and an inhaled anticholinergic bronchodilator can slow down the decline in forced expired volume in 1 s ($FEV_1$) during a 5-year follow-up period. Patients' $FEV_1$ values were collected annually upon recruitment into the study. $FEV_1$ was chosen as the primary outcome since its trajectory is an indicator of a patient's natural loss of lung function during the progression of COPD. Since not all patients completed the whole study period, about 9.47% of longitudinal measurements were missing. One of the possible reasons for dropout is that treatment was not effective, and hence, missing longitudinal measurements after dropout are nonignorable.

Joint modeling of $FEV_1$ together with the possible informative dropout time provides an attractive approach to deal with nonignorable missing longitudinal data due to dropout. Based on previous findings, we considered the following

Table 1: Runtime comparison between different R packages for joint modeling of a longitudinal and a single event time on the lung health study data.

| Package | Semiparametric joint models | | | | | Parametric joint models | | | |
| --- | --- | --- | --- | --- | --- | --- | --- | --- | --- |
| | FastJM | joineR | $JSM_a$ | $JSM_b$ | $JM_b$ | $JM_{a_1}$ | $JM_{a_2}$ | $JMBayes_a$ | $JMBayes_b^*$ |
| Baseline hazard | Unspecified | Unspecified | Unspecified | Unspecified | Unspecified | Weibull | B-spline | B-spline | B-spline |
| $W_i(t)$ | $v^T b_i$ | $v\tilde{X}_i^{(1)}(t)^T b_i$ | $v m_i(t)$ | $v\tilde{X}_i^{(1)}(t)^T b_i$ | $v m_i(t)$ | $v m_i(t)$ | $v m_i(t)$ | $v m_i(t)$ | $v^T b_i$ |
| Runtime | 0.3 min | 20.4 min | 1 h 36 min | 1 h 51 min | * | 0.9 min | 1 min | 19.8 min | 43 min |

*Failed to produce any result due to convergence issue.

covariates when characterizing the trajectory of $Y = FEV_1$: time (year), sex, age, body mass index (BMI), baseline number of cigarettes smoked per day, and the logarithm of two-point methacholine concentration-$FEV_1$ O'Connor slope (logslope) [31]. We also included two interaction terms between treatment indicators SIP and SIA and time, so that the difference in the slope of $FEV_1$ between SIP (or SIA) and usual care can be evaluated by testing if the interactions are zero or not. Specifically, we considered the following linear mixed effects model:

$$Y_{ij} = \beta_0 + \beta_1 t_{ij} + \beta_2 X_{age_i} + \beta_3 X_{F10CIGS_i} + \beta_4 D_{sex_i} + \beta_5 X_{logslope_i} \\ + \beta_6 X_{BMI_i} + \beta_7 D_{SIP_i} + \beta_8 D_{SIA_i} + \beta_9 D_{SIP_i} \times t_{ij} + \beta_{10} D_{SIA_i} \\ \times t_{ij} + b_{0i} + b_{1i} t_{ij} + \varepsilon_{ij},$$

(26)

which corresponds to model (1) with $X_i^{(1)}(t_{ij})^T = (1, t_{ij}, X_{age_i}, X_{F10CIGS_i}, D_{sex_i}, X_{logslope_i}, X_{BMI_i}, D_{SIP_i}, D_{SIA_i}, D_{SIP_i} \times t_{ij}, D_{SIA_i} \times t_{ij})$ and $\tilde{X}_i^{(1)}(t_{ij})^T = (1, t_{ij})$. The random error term $\varepsilon_{ij} \sim^{iid} N(0, \sigma^2)$ and the random effects $b_i = (b_{0i}, b_{1i})^T$ are assumed normally distributed with zero mean and a covariance matrix $\Sigma$. For the dropout time $T_i$ (possibly censored at the end of the study), we assume the Cox proportional hazard submodel.

$$\lambda_i(t) = \lambda_0(t) \exp\{\gamma_1 X_{BMI_i} + \gamma_2 D_{SIP_i} + \gamma_3 D_{SIA_i} + \gamma_4 X_{logslope_i} \\ + \gamma_5 D_{sex_i} + \gamma_5 X_{age_i} + W_i(t)\},$$

(27)

where $\lambda_0(t)$ denotes the baseline hazard function and $W_i(t)$ is a latent association structure that links the two submodels.

Table 1 compares the runtime of FastJM and some existing joint model packages including joineR [29], different versions of JM [2], JMbayes [32], and JSM [33] with various specifications of $\lambda_0(t)$ and $W_i(t)$.

Among all the semiparametric models (FastJM, joineR, $JSM_a$, and $JSM_b$), FastJM finished in 0.3 minutes while other methods took 20.4 minutes to 111 minutes. As a matter of fact, the runtime of FastJM was even shorter than those of some parametric joint models ($JSM_a$ and $JSM_b$). We also observed that JMbayes based on a Bayesian MCMC framework is considerably slower than its frequentist counterpart JM. Finally, the parameter estimates and inference results for the longitudinal outcome were almost identical between all packages, but slightly different for the survival submodel because of their slightly different latent structure $W_i(t)$. Detailed analysis results are summarized in Section A.7 of the supplementary materials.

4.2. UK Biobank Primary Care (UKB-PC) Study. The UK Biobank (UKB) is a prospective cohort study with deep genetic and phenotypic data collected on approximately 500,000 individuals, aged 37-73 years, from the general population between 2006 and 2010 in the United Kingdom [18, 19]. Participants attended assessment at their closest clinic center where they completed questionnaires, took physical measurements, and provided biological samples (blood, urine, and saliva) as a baseline assessment visit. Hospital admission records were available until February, 2018, for the full UKB cohort, whereas linkage to primary care records was available for 45% of the UKB cohort (approximately 230,000 participants) until May, 2017, for Scotland, September, 2017, for Wales, and August, 2017, for England. The detailed linkage procedures relating to primary care records are available online (https://biobank.ndph.ox.ac.uk/showcase/showcase/docs/primary_care_data.pdf).

In this example, we consider a joint model of longitudinal systolic blood pressure (SBP) measurements and a competing risk event time defined as age-at-onset of type 2 diabetes (T2D) as the first risk and age-at-onset of stroke, myocardial infarction (MI), or all-cause death as the second risk, whichever occurred first. Age-at-onset of outcomes were based on participants' primary care or hospital records, whichever occurred first. Follow-up was censored at the primary care data end date for the relevant country or the date of outcomes, if this occurred earlier. SBP measures were extracted from either baseline assessment visit or primary care data. Covariates include sex, ethnicity, and BMI measured during baseline visit. However, considering the imbalanced racial distribution in this case study, we only considered white vs. non-white ethnicity groups. Specifically, the joint model consists of a linear mixed effects model for the longitudinal outcome (SBP),

$$Y_{ij} = \beta_0 + \beta_1 t_{ij} + \beta_2 X_{BMI_i} + \beta_3 D_{male_i} + \beta_4 D_{non-white_i} + b_{0i} \\ + b_{1i} t_{ij} + \varepsilon_{ij}, \text{(Model} - L)$$

(28)

which corresponds to model (1) with $X_i^{(1)}(t_{ij})^T = (1, t_{ij}, X_{BMI_i}, D_{male_i}, D_{non-white_i})$ and $\tilde{X}_i^{(1)}(t_{ij})^T = (1, t_{ij})$ and a proportional cause-specific hazard model for the competing risk event outcome

$$\lambda_{ik}(t) = \lambda_{k0}(t) \exp \{\gamma_{k1} X_{\text{BMI}_i} + \gamma_{k2} D_{\text{male}_i} + \gamma_{k3} D_{\text{non-white}_i} \\ + W_{ik}(t)\}, \text{(Model – PCH)} \quad (29)$$

for $k = 1, 2$. In Model-L, the random error term $\varepsilon_{ij} \sim^{\text{iid}} N(0, \sigma^2)$ and the random effects $b_i = (b_{0i}, b_{1i})^T$ are assumed normally distributed with zero mean and covariance matrix $\Sigma$. In Model-PCH, $k = 1$ denotes type 2 diabetes and $k = 2$ stroke, $\lambda_{k0}(t)$ denotes the baseline cause-specific hazard function for cause $k$, and $W_{ik}(t)$ denotes the latent association structure of SBP with cause $k$ risk.

To our knowledge, besides our FastJM package, joineR and JM are two other current joint model R packages that are capable of handling competing risk event outcomes. However, because JM encountered convergence issues, we will focus on FastJM and joineR in this case study. Table 2 compares the runtime of FastJM and joineR on a subset of 5,000 and 20,000 participants randomly selected from the UKB-PC data and the full UKB-PC data with 193,287 participants.

Table 2 shows that for the UKB-PC subset of 5,000 participants, FastJM finished within 1 minute, while joineR took 3.3 hours to finish. For the UKB-PC subset of 20,000 participants, FastJM finished within 5 minutes, while joineR took 33 hours to run. For the UKB-PC full data with 193,287 participants, FastJM finished within 1 hour, whereas joineR encountered a computational failure.

Finally, the analysis results produced by FastJM and joineR are similar for the longitudinal submodel for the UKB-PC subset of 5,000 and 20,000 participants and for UKB-PC full data. For the survival submodel, the analysis results are also similar for most parameters except for the association parameters due to the different latent structure $W_i(t)$ between two packages. Detailed analysis results are provided in Section A.8 of the supplementary materials.

## 5. Discussion

We have developed customized linear scan algorithms to reduce the computational complexity from $O(n^2)$ or $O(n^3)$ to $O(n)$ within each iteration of the EM algorithm and in the standard error estimation step for a semiparametric joint model of a longitudinal biomarker and a competing risk event time outcome. Through simulation and case studies, we have demonstrated that the efficient implementation can generate a speed-up by a factor of up to hundreds of thousands and oftentimes reduce the runtime from days to minutes when the sample size is large ($n > 10^4$), making it feasible to fit a joint model on a large data in real time.

The ideas and techniques of this paper can potentially be adapted to improve computational efficiency for other joint models. For instance, the linear computational algorithm in Remark 1 for computing the variance-covariance matrices of empirical Bayes estimates of the random effects is not spe-

TABLE 2: Runtime comparison between different R packages for semiparametric joint modeling of longitudinal SBP trajectory and competing risk event time on the UK Biobank primary care (UKB-PC) data.

| Package | FastJM | joineR |
| --- | --- | --- |
| UKB-PC subset ($n = 5,000$) | 1 min | 3.3 h |
| UKB-PC subset ($n = 20,000$) | 4.4 min | 33 h |
| UKB-PC full data ($n = 193,287$) | 1 h | * |

*Failed to produce any result due to computational failure.

cific to the joint model considered in this paper and can be used in any procedure that uses the pseudo-adaptive quadrature rule. Also, although we have focused on joint modeling of a single biomarker with a time-to-event outcome, our methodology can be easily extended to handle multiple biomarkers in a similar fashion. It is also important to note that the linear risk set scan algorithm is limited to the share random effect joint model in which the Cox submodel (2) only involves time-independent covariates. If the Cox submodel contains time-dependent covariates such as the present value of the longitudinal marker, then one may have to impose more restrictive assumptions such as assuming a parametric baseline hazard in order to linearize the computation costs with respect to the sample size.

This paper has focused on linearizing the computation with respect to the sample size within the framework of classical EM algorithm that is coupled with the pseudo-adaptive quadrature rule for numerical integration in the E-step. It would be interesting to investigate if coupling our algorithms with other numerical integration methods such as quasi-Monte Carlo method [34] in the E-step or with other variations of EM algorithms such as the stochastic EM algorithm (stEM) [35] or Turbo-EM [36] could further enhance the computational efficiency, especially when there are 3 or more random effects in the model. Finally, current joint model implementations are generally not scalable as the number of longitudinal measurement grows large, rendering it infeasible to fit dense longitudinal data such as those generated from modern wearable devices for dynamic prediction of a health outcome. Future research is warranted to develop novel joint modeling procedures that are scalable to large number of subjects, random effects, and longitudinal measurements.

## 6. Software

A user-friendly R package FastJM [37] has been developed to fit the shared parameter joint model using the efficient algorithms developed in this paper and is publicly available on the Comprehensive R Archive Network (CRAN) at https://CRAN.R-project.org/package=FastJM.

## Data Availability

The lung health study (LHS) data used to support the findings in Section 4.1 have not been made available because the authors did not have permission for data sharing from the data provider. The data that support the findings in Section 4.2 are available from UK Biobank repositories. The UK Biobank data are retrieved under Project ID: 48152. Data are available at https://www.ukbiobank.ac.uk with the permission of UK Biobank.

### Disclosure

This paper has been submitted as preprint as per the URL: https://arxiv.org/abs/2110.14822 [38].

### Conflicts of Interest

The authors declare that they have no conflicts of interest.

### Acknowledgments

This research was partially supported by the National Institutes of Health (P30 CA-16042, UL1TR000124-02, and P01AT003960, GL), the National Institute of General Medical Sciences (R35GM141798, HZ), the National Human Genome Research Institute (R01HG006139, HZ and JJZ), the National Science Foundation (DMS-2054253, HZ and JJZ), the National Institute of Diabetes and Digestive and Kidney Disease (K01DK106116, JJZ), and the National Heart, Lung, and Blood Institute (R21HL150374, JJZ).

### Supplementary Materials

Derivations of formulas, additional simulation results, and analysis results of the two real data are provided in the supplementary materials. *(Supplementary Materials)*